\documentclass[aps,preprintnumbers,superscriptaddress,nofootinbib,amsmath,amssymb,amsfonts,showpacs,floatfix]{revtex4-1}

\usepackage{epstopdf}
\usepackage{afterpage}
\usepackage{mathrsfs}

\usepackage{bm}
\usepackage{setspace}

\usepackage{mathrsfs}
\usepackage{amsfonts}
\usepackage{amsmath}
\usepackage{bm}
\usepackage{array}
\usepackage{verbatim}
\usepackage{epsfig}
\usepackage{graphicx} 
\usepackage[svgnames]{xcolor}
\usepackage{float}
\usepackage{hyperref}

\usepackage{xspace}

\def\bea#1\eea{\begin{align}#1\end{align}}

\begin{document}

 \title{Pre-Town Meeting on Spin Physics at an Electron-Ion Collider\footnote{Mini--review summarizing the Informal Pre-Town Meeting at Jefferson Lab, {\bf{http://www.jlab.org/conferences/pretownjlab2014/}}}}

\author{Elke-Caroline Aschenauer}
\affiliation{Physics Department, Brookhaven National Laboratory, Upton, New York 11973, USA}

\author{Ian Balitsky}
\affiliation{Physics Department, Old Dominion University, Norfolk, Virginia 23529, USA}
\affiliation{Jefferson Lab, 12000 Jefferson Avenue, 
Newport News, Virginia 23606, USA}

\author{Leslie Bland}
\affiliation{Physics Department, Brookhaven National Laboratory, Upton, New York 11973, USA}

\author{Stanley J.  Brodsky}
\affiliation{SLAC National Accelerator Laboratory, Stanford University, Stanford, California94309, USA}

\author{Matthias Burkardt}
\affiliation{Department of Physics, New Mexico State University, Las 
Cruces, NM 88003-8001, USA}

\author{Volker Burkert} 
\affiliation{Jefferson Lab, 12000 Jefferson Avenue, Newport News, Virginia 23606, USA}

\author{Jian-Ping Chen}
\affiliation{Jefferson Lab, 12000 Jefferson Avenue,  Newport News, Virginia 23606, USA}

\author{Abhay Deshpande} 
\affiliation{RIKEN BNL Research Center, Brookhaven National Laboratory, Upton, New York 11973-5000, USA}
\affiliation{Department of Physics and Astronomy, Stony Brook University, SUNY, Stony Brook, New York 11794-3800, USA}

\author{Markus Diehl}
\affiliation{Deutsches Elektronen-Synchroton DESY, 22607 Hamburg, Germany}

\author{Leonard Gamberg}
\email{lpg10@psu.edu}
\affiliation{Division of Science, 
Penn State University-Berks, Reading, PA 19610, USA}

\author{Matthias Grosse Perdekamp}
\affiliation{University of Illinois at Urbana-Champaign, Urbana, Illinois 61801, USA}

\author{Jin Huang}
\affiliation{Massachusetts Institute of Technology, Cambridge, Massachusetts 02139}

\author{Charles Hyde}
\affiliation{Physics Department, Old Dominion University, Norfolk, Virginia 23529, USA}

\author{Xiangdong Ji}
\affiliation{INPAC, Department of Physics, and Shanghai Key Lab for Particle Physics and Cosmology, Shanghai Jiao Tong University, Shanghai, 200240, P. R. China}
\affiliation{Center for High-Energy Physics, Peking University, Beijing, 100080, P. R. China}
\affiliation{Maryland Center for Fundamental Physics,
University of Maryland, College Park, Maryland 20742, USA}

\author{Xiaodong Jiang}
\affiliation{Los Alamos National Laboratory, Los Alamos, New Mexico 87545, USA}

\author{Zhong-Bo Kang}
\email{zkang@lanl.gov}
\affiliation{Theoretical Division, Los Alamos National Laboratory, 
Los Alamos, New Mexico 87545, USA}

\author{Valery Kubarovsky}
\affiliation{Jefferson Lab, 12000 Jefferson Avenue,  Newport News, Virginia 23606, USA}

\author{John Lajoie}
\affiliation{Iowa State University, Ames, Iowa 50011, USA}

\author{Keh-Fei Liu} 
\affiliation{Dept. of Physics and Astronomy
Center for Computational Sciences
University of Kentucky, Lexington, Kentucky 40506, USA}

\author{Ming Liu}
\affiliation{Los Alamos National Laboratory, Los Alamos, New Mexico 87545, USA}

\author{Simonetta Liuti}
\affiliation{Department of Physics, University of Virginia, Charlottesville, Virginia 22904, USA}

\author{Wally Melnitchouk}
\affiliation{Jefferson Lab, 12000 Jefferson Avenue, Newport News, Virginia 23606, USA}

\author{Piet Mulders}
\affiliation{Nikhef and Department of Physics and Astronomy, VU University Amsterdam,
De Boelelaan 1081, NL-1081 HV Amsterdam, the Netherlands}

\author{Alexei Prokudin}
\email{prokudin@jlab.org}
\affiliation{Jefferson Lab,  12000 Jefferson Avenue,  Newport News, Virginia 23606, USA}

\author{Andrey Tarasov}
\affiliation{Jefferson Lab,  12000 Jefferson Avenue,  Newport News, Virginia 23606, USA}

\author{Jian-Wei Qiu}
\affiliation{Physics Department, Brookhaven National Laboratory, Upton, NY 11973, USA}
\affiliation{C.N. Yang Institute for Theoretical Physics and Department
of Physics and Astronomy,
Stony Brook University, Stony Brook, NY 11794-3840, USA}

\author{Anatoly Radyushkin}
\affiliation{Physics Department, Old Dominion University, Norfolk, Virginia 23529, USA}
\affiliation{Jefferson Lab, 12000 Jefferson Avenue, 
Newport News, Virginia 23606, USA}

\author{David Richards}
\affiliation{Jefferson Lab, 12000 Jefferson Avenue, Newport News, Virginia 23606, USA}

\author{Ernst Sichtermann}
\affiliation{Nuclear Science Division, 
Lawrence Berkeley National Laboratory, Berkeley, California 94720, USA}

\author{Marco Stratmann}
\affiliation{Institute for Theoretical Physics, T\"ubingen University, Auf der Morgenstelle 14, 72076 T\"ubingen, Germany}

\author{Werner Vogelsang} 
\affiliation{Institute for Theoretical Physics, T\"ubingen University, Auf der Morgenstelle 14, 72076 T\"ubingen, Germany}

\author{Feng Yuan}
\affiliation{Nuclear Science Division, 
Lawrence Berkeley National Laboratory, Berkeley, California 94720, USA}

\begin{abstract}
A polarized $ep/eA$ collider (Electron--Ion Collider, or EIC), with polarized proton and light-ion beams and unpolarized heavy-ion beams with a variable
center--of--mass energy $\sqrt{s}  \sim 20$ to $\sim100$~GeV (upgradable to $\sim 150$ GeV) and a luminosity 
up to $\sim 10^{34} \, \textrm{cm}^{-2} \textrm{s}^{-1}$, would be uniquely
suited to address several outstanding questions of Quantum Chromodynamics, and 
thereby lead to new qualitative and quantitative information
on the  microscopic structure of hadrons and nuclei. During this meeting at 
Jefferson Lab we addressed recent theoretical and experimental  
 developments in the spin and the 
three--dimensional structure of the nucleon  (sea quark and gluon
spatial distributions, orbital motion, polarization, and their correlations).  
This mini--review contains a short update on progress in these areas since the 
 EIC White paper~\cite{Accardi:2012qut}.  
\end{abstract}
\date{\today}

 \maketitle

%
%
\newpage
\begin{flushleft}
{\bf Introduction:}
\end{flushleft}
A group of 44 physicists from the spin physics community gathered at Jefferson Lab from August 12-15, 2014 to discuss the current status of an 
Electron--Ion Collider (EIC) and its impact on the future development of nuclear physics, and in particular on 
spin physics and the three-dimensional (``3-D'') structure of hadrons (with a 
 focus on nucleon structure).

Participants were asked to prepare short presentations concerning the important scientific questions in spin physics and in particular in ``3-D'' nucleon structure which could be addressed and answered by an 
EIC. In order to facilitate discussions, the organizers appointed ``advocates'' and ``convener-opponents''
for each topic discussed. The  ``advocates'' were given time to present their view of future impact of EIC physics, while ``convener-opponents'' were asked to lead and direct discussions,  posing in depth questions about 
the  content and justifications presented by ``advocates''.
The discussions were followed by short presentations with an opportunity for
 ample contributions from all participants. In particular the discussions were intended to summarize developments in both theory and experiment in the last two years that followed the completion of the EIC White Paper~\cite{Accardi:2012qut}.

This  document summarizes the progress discussed during the ``Informal Pre-town Meeting at Jefferson Lab", August 12-15, 2014.
All presented talks can be found on the workshop website: \\
\url{http://www.jlab.org/conferences/pretownjlab2014/program.html}.\\

\begin{flushleft}
{\bf General discussion:}
\end{flushleft}
As a basic building block of more than 99\% of the mass of the visible world, the nucleon is a confined system of strongly interacting quarks, anti-quarks, and gluons (partons). Understanding  this complex structure and the nature of the  strong force is a great scientific challenge and one of  the paramount problems of modern nuclear physics, as 
described in the 2007 NSAC Long--Range Plan~\cite{NSACpref}. It is an essential step in describing nuclear structure and reactions
from first principles, with numerous applications to science and technology. The past century has been marked by a truly remarkable achievement: 
Quantum Chromodynamics (QCD), the fundamental theory of the strong force, has been tested and established. Experiment has yielded precision insights on the proton's spatial extent, on the one hand, and its momentum structure on the other, 
revealed the existence of a large number of soft gluons, and resulted in  
 many  surprises  with  nuclei and spin. 
 The ongoing and planned programs at Jefferson Lab, RHIC and other facilities 
in Europe and Japan will allow us to significantly extend these studies and for the first time gain precision knowledge of the combined spatial-momentum structure of the nucleon in the region dominated by valence quarks,  as well as expanding our 
understanding of the partonic sea of nucleons and nuclei.

A polarized $ep/eA$ Electron-Ion Collider  with
 polarized proton and light-ion beams and unpolarized heavy-ion beams,
 with a variable $ep$ center--of--mass (CM) energy in the range
 $\sqrt{s} \sim 20$ to $\sim 100 \, \textrm{GeV}$ (upgradable to
 $\sim 150$ GeV), will be a uniquely powerful ``femtoscope" to improve
 our understanding of the nucleon and the nucleus in the ``sea" quark
 and gluon-dominated regime at high $Q^2$.  With its unique
 versatility and state-of-the-art luminosity of up to
 $\sim 10^{34} \, \textrm{cm}^{-2} \textrm{s}^{-1}$,
 its high polarization, and detector capabilities
 (as specified in the EIC White Paper~\cite{Accardi:2012qut}
 and Refs.~\cite{Boer:2011fh,Anselmino:2011ay,Accardi:2011mz}),
 it will achieve a precision far beyond that of any other existing
 and planned facilities in the world.

The inability to observe quarks and gluons in isolation, owing to QCD confinement, presents a great challenge in studying the interior structure of
nucleons and nuclei. An EIC with its unprecedented precision may provide data
which help to address this challenge by probing nucleon structure with
 increasing accuracy.  The combination of precision data from a broad range of reactions at an EIC,  with commensurate advances in the theory 
of QCD dynamics of quarks, anti-quarks, and gluons,  will lead to the discovery of  new phenomena and will reveal new structures inside hadrons. In turn this will  yield unprecedented understanding of nucleon and nuclear properties in terms of these elementary constituents
and their dynamics, as well as the mechanism for the emergence of hadrons. Theoretical methods to apply QCD to hadronic and nuclear systems have 
made dramatic advances in the last two decades, but rely crucially
on new experimental information for further progress.

An example is the gluon spin: An EIC can delineate with unprecedented 
precision,  over an extended kinematic region in Bjorken-$x$,  the full helicity structure of the proton in terms of gluons, quarks, and anti-quarks and their flavor. This will address one very important aspect of the nucleon spin puzzle.

Another example is the combined spatial and momentum  parton structure in the nucleon in terms of densities, which  are encoded in the transverse-momentum-dependent distributions (TMDs) and generalized parton distributions (GPDs).  In addition to providing 
a tomographic description of the nucleon, the precision mapping of TMDs and GPDs will provide new qualitative and quantitative information about the orbital motion of the partons and, with further development of phenomenological models, and ab initio lattice calculations, quantitative information about orbital angular momentum.

%
%
\begin{figure}[!htb]
  \centering
  \parbox{0.40\textwidth}{\includegraphics[width=0.4\textwidth]{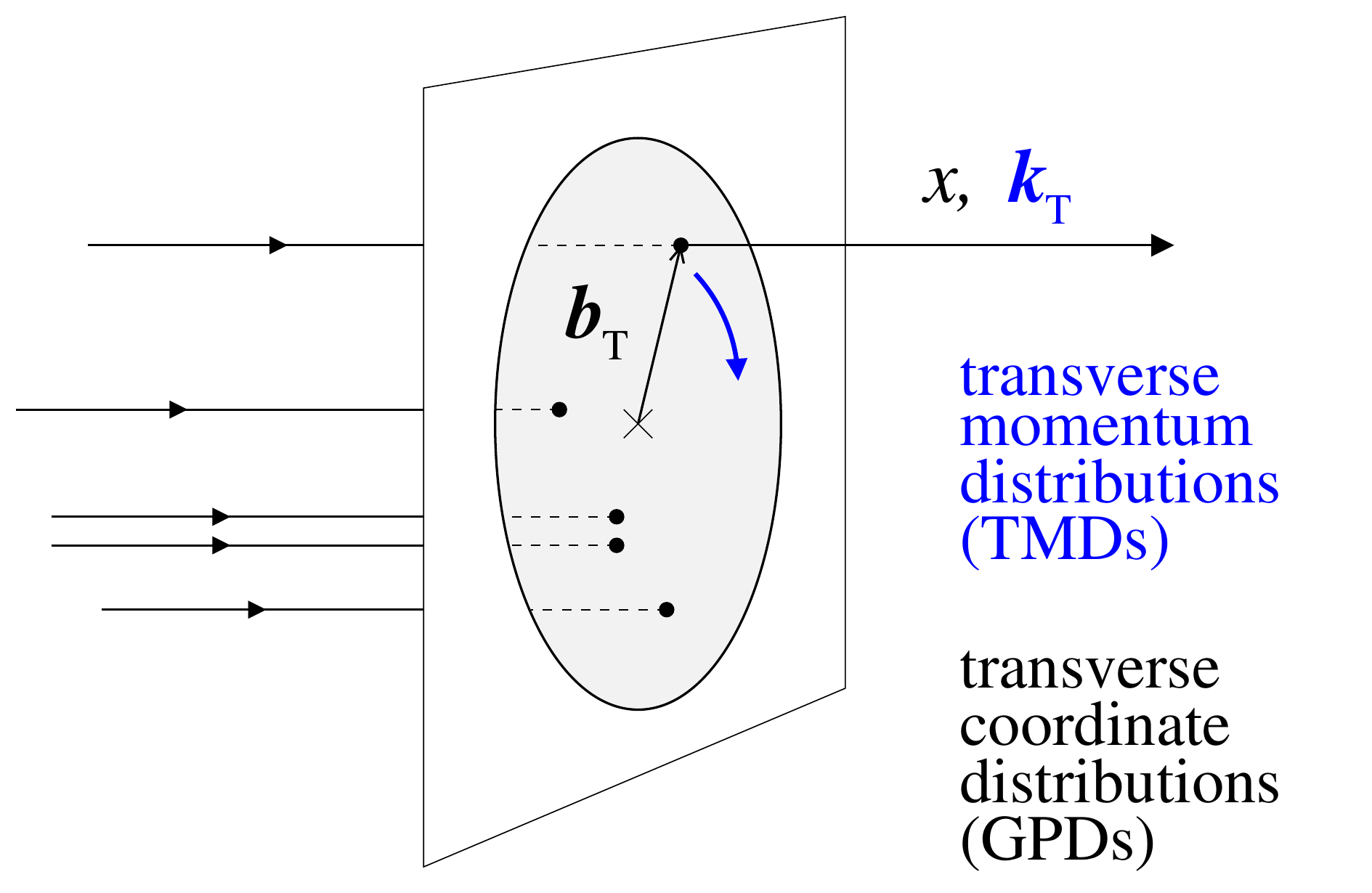}}
  \hspace{0.04\textwidth}
  \parbox{0.54\textwidth}{\caption{
Three--dimensional structure of a fast--moving 
nucleon. The distribution of partons (quarks, gluons) is
characterized by the longitudinal momentum fraction $x$ and the
transverse spatial coordinate $\bm{b}_T$ 
through the impact parameter GPDs~\cite{Burkardt:2000za,*Burkardt:2002hr}. In addition, the 
partons are distributed over transverse momenta $\bm{k}_T$, 
reflecting their orbital motion and interactions in the system (TMDs). 
Polarization distorts both the spatial and momentum distributions. 
Note that $\bm{b}_T$ and $\bm{k}_T$ are not 
Fourier conjugate; a joint description in both variables can be 
formulated in terms of a Wigner phase space density~\cite{Ji:2003ak}.
Observables sensitive to either $\bm{b}_T$ or $\bm{k}_T$ help to 
establish a three--dimensional dynamical picture of the nucleon in QCD. Figure from Ref.~\cite{Accardi:2011mz}.}
\label{fig:TMD}
}
\end{figure}
\newpage
\begin{flushleft}
{\bf Three--dimensional structure of the nucleon in QCD:}
\end{flushleft}
The nucleon in QCD represents a dynamical system of fascinating complexity. 
Viewed in high--energy interactions the nucleon's color field can be
represented by elementary quanta with  point--particle characteristics (partons), and the nucleon becomes a 
many--body system of quarks, anti-quarks and gluons.  In contrast 
to   non--relativistic systems, in QCD the number of   
constituents is not fixed, as they constantly undergo creation/annihilation 
processes mediated by QCD interactions. This reflects the 
relativistic nature of the dynamics. A
 high--energy scattering process 
takes a ``snapshot'' of this fast--moving system with a spatial resolution 
given by the inverse momentum transfer $1/Q$.  
In addition to the valence quarks, the nucleon contains 
a ``sea'' of quark--anti-quark pairs.  The spin and 
flavor quantum numbers carried by the sea, both light and heavy, are poorly constrained by present data.  An EIC would be  the only polarized ``femtoscope" facility in the world that
is capable of accessing distances $< 10^{-15}$ cm inside of the polarized nucleon in the regime where sea quarks and gluons dominate.

Experimental measurements of semi-inclusive deep inelastic scattering (SIDIS) and exclusive processes such as 
deeply virtual Compton scattering (DVCS) lead to an understanding of a ``three--dimensional''  representation of the nucleon in coordinate and momentum space. New distributions related to those representations are
so-called Generalized Parton Distributions   and Transverse Momentum Dependent distributions, see Fig.~\ref{fig:TMD}. 
A  joint description in both momentum and coordinate space can be 
formulated in terms of Wigner phase space densities~\cite{Ji:2003ak}, however, it remains a  challenge to find a way of accessing 
these distributions in high-energy scattering
experiments. Measurements with polarized beams of light ions  and electrons are key requirements for understanding GPDs and TMDs.  An EIC will measure TMDs of sea quarks through semi--inclusive measurements, 
in which the charge and flavor of the struck quark/anti-quark are ``tagged'' 
by detecting hadrons ($\pi^\pm, K^\pm, p, \bar p, \ldots)$ produced from 
its fragmentation.

Since  the publication of the EIC White Paper~\cite{Accardi:2012qut}, theoretical tools for QCD evolution of TMDs have been developed and have been successfully implemented in phenomenological calculations (see contributions in~\cite{Balitsky:2011jrs,*Radyushkin:2012doa,*Radyushkin:2014kla}). This progress allows one to cover much larger energy  momentum ranges, and thereby implement QCD-based fits that are able to describe data coming from low to medium energies, such as Jefferson Lab 6, HERMES, COMPASS up to
high energies of LHC.  Recent progress  in SIDIS experiments,
 and publication of unpolarized multiplicities from  Jefferson Lab, HERMES, and COMPASS 
allowed for the first time successful implementation of QCD evolution of TMDs.  Future measurements of the Drell-Yan process at COMPASS, FERMILAB, and RHIC 
will   probe for the first time the universality
 properties   of those distributions~\cite{Gautheron:2010wva,Aidala:2012nz,Star:2014,drellyan:2014f}. 
In its report to NSAC by the subcommittee on performance measures 
this was identified
as one of the high priority milestones (HP-13)~\cite{NSACpref}: Namely, to 
``Test unique QCD predictions for
relations between single-transverse spin phenomena in $pp$ scattering
and those observed in deep-inelastic lepton scattering''. 
However, a truly quantitative measurement of TMDs will require the
    large multi-dimensional phase space and the high luminosity of an EIC,
    and only at this facility will we be able to explore the region of sea
    quarks.  EIC kinematics is ideally suited to probe the interplay
    between non perturbative and perturbative dynamics in generating
    transverse momentum and to quantify the non perturbative contribution
    to the scale evolution of TMDs. 
 The data from an EIC will be crucial for future progress in the development of the theory and phenomenology of TMDs as it relates to the ``3-D'' momentum structure of the nucleon and nuclei.

GPDs define the basic size and shape of the nucleon in QCD,   
 generalizing the  one--dimensional picture conveyed from  the longitudinal 
momentum  densities into a ``two spatial  plus one momentum''--dimensional 
image of the fast moving 
nucleon \cite{Burkardt:2000za,*Burkardt:2002hr}. 
Information on the transverse distribution of quarks and gluons
is obtained from exclusive scattering $\gamma^\ast N \rightarrow M + N
\, (M = \textrm{meson}, \gamma, \textrm{heavy quarkonium})$. 
 GPDs 
combine the concept of the quark/gluon momentum density with that of
elastic nucleon form factors. Measurements of $J/\psi$ photo/electro-production
with an EIC provide a unique opportunity
 to map the transverse spatial 
distribution of gluons in the nucleon above 
$x \sim \textrm{few} \times 10^{-3}$ in unprecedented detail.

%
%

\begin{figure}[t]
  \centering
  \parbox{0.65\textwidth}{\includegraphics[width=0.69\textwidth]{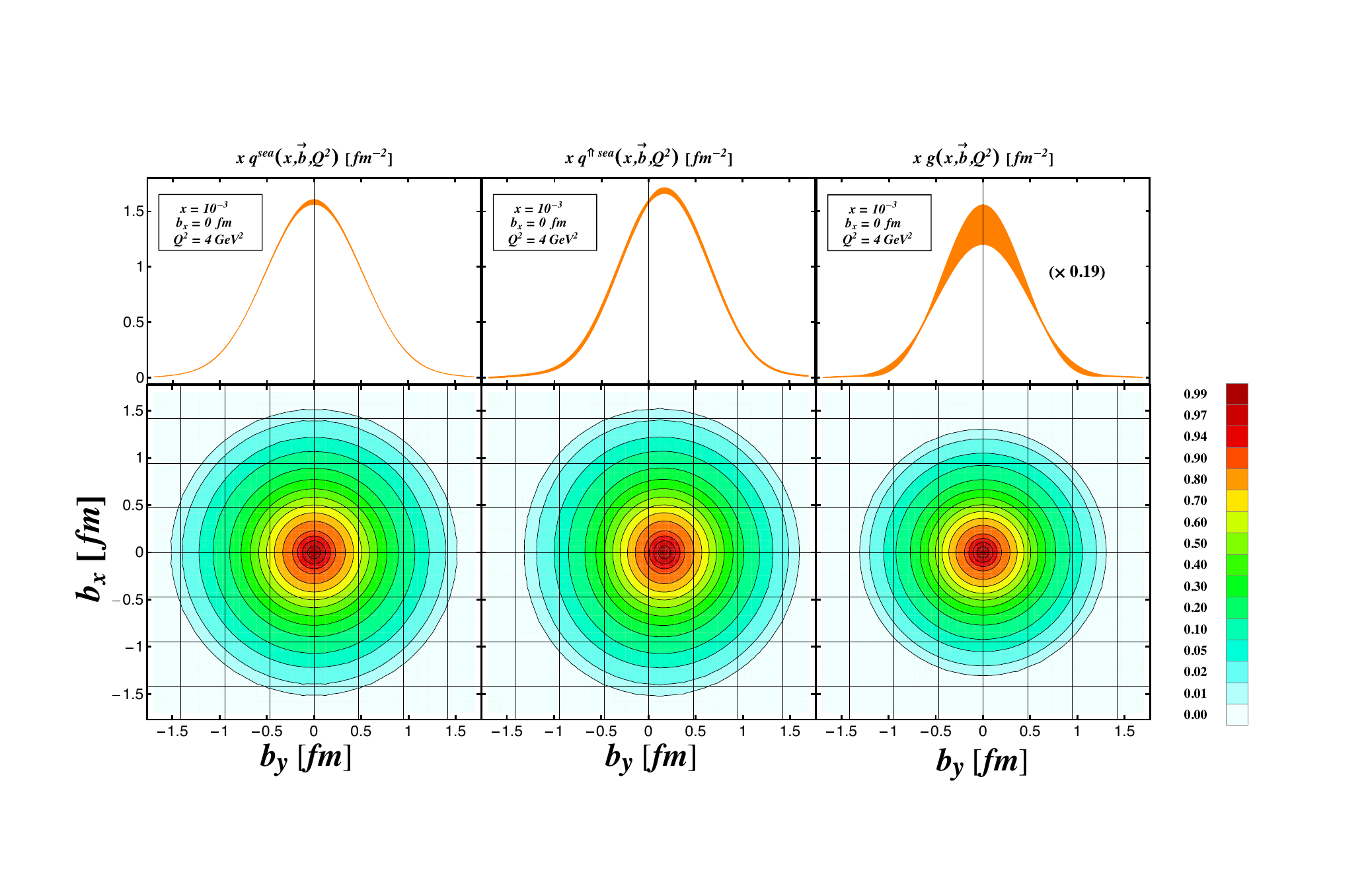}}
  (a) \\
  \parbox{0.65\textwidth}{\includegraphics[width=0.55\textwidth]{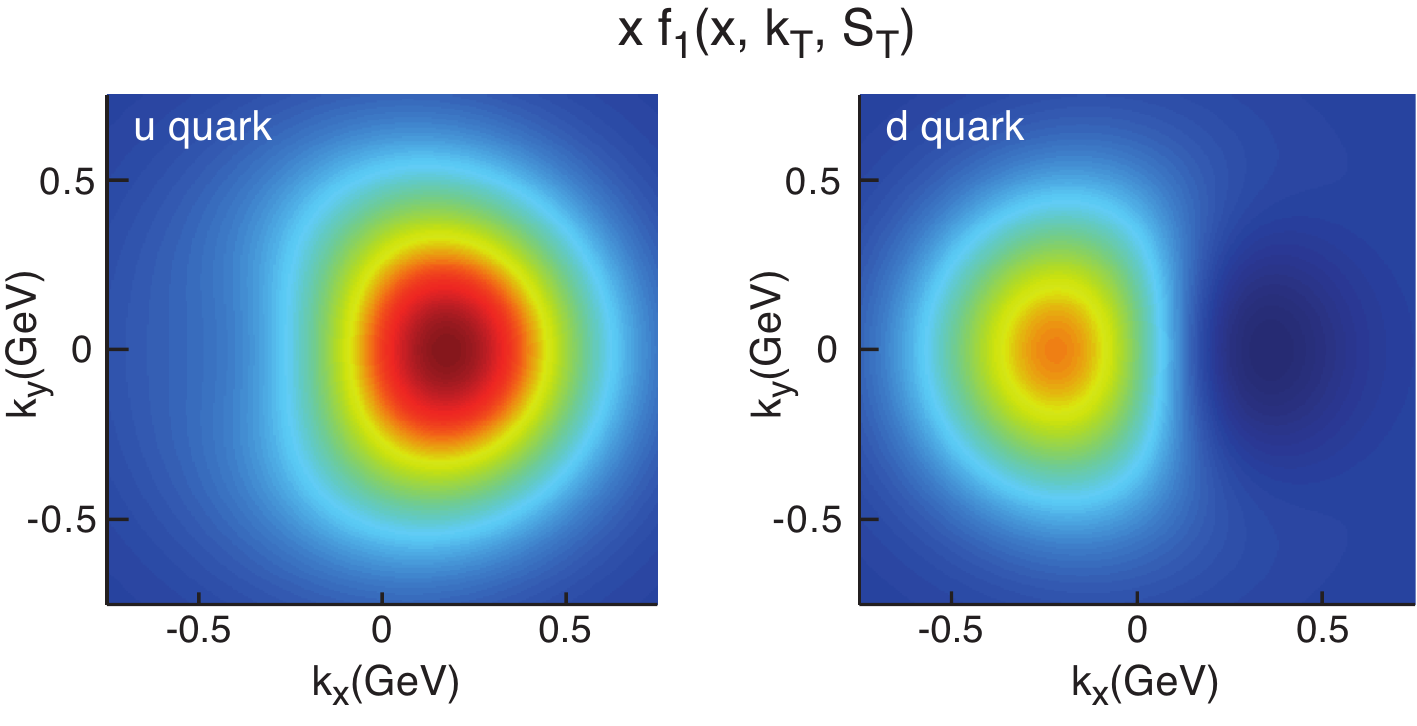}} (b) 
  \hspace{0.04\textwidth}
  \parbox{1.\textwidth}{\caption{
Examples of three--dimensional structure: (a) 
Parton densities at $x = 0.001$ and $Q^2 = 4 \mathrm{GeV}^2$ 
versus impact parameter $b$  obtained from a combined least-squares fit to 
HERA collider and EIC pseudo-data: relative densities (lower row) and their values at $b_x = 0$ 
for the unpolarized sea quark parton densities of a unpolarized proton (left), a transversely polarized proton (middle), and the unpolarized gluon parton density of a unpolarized proton (right), its value is re-scaled by a factor $0.19$.
 Figure from Ref.~\cite{Aschenauer:2013hhw}.
(b) current knowledge of TMD distributions for $u$ and $d$ quarks at $x=0.1$ as function of $k_x$ and $k_y$ presented as ``tomographic" slice. Figure from Ref.~\cite{Accardi:2012qut}.}
\label{fig:3D}
}
\end{figure}

The EIC will map the spatial distributions of quarks and gluons over a wide range of the longitudinal momentum fraction.
We expect large differences in these spatial distributions for the 
charge, quark-matter, and gluon-matter distributions in the region of 
$x < 0.1$. 

In the valence region ($x>0.1$) the EIC will add a large lever arm 
in $Q^2$ to complement the data  obtained from the  
Jefferson Lab and COMPASS programs.  The up-, down-, and 
strange-quark distributions do not follow a simple scaling based on
their current quark masses, and there is evidence from PDF fits
to the HERA DIS data that there is significant strength in the gluon
distributions at high $x$.  In addition to the sea, 
generated by gluon-splitting, 
QCD predicts from first principles the existence of heavy quarks 
at large $x$ in the proton wave function.  Such distributions are crucial 
for understanding the production of  hadrons containing heavy quarks both 
at threshold and at high longitudinal momentum. Direct evidence for such ``intrinsic" heavy quarks has been difficult to obtain from existing data \cite{Jimenez-Delgado:2014zga}, and an EIC has the potential to discover this component, if it exists at non-negligible levels.

There is a lot of activity in the calculation of TMDs and GPDs 
in various non-perturbative models, such as the Dyson-Schwinger approach, 
holography and AdS/QCD~\cite{deTeramond:2008ht,*Brodsky:2014yha}.  In the Hamiltonian approach these 
distributions are related to eigen-solutions 
of the QCD light-front Hamiltonian.  The distributions  of the nucleon 
which can be measured at an  EIC can be constructed from 
operators in lattice QCD calculations. GPDs can be constructed from 
local operators and  lattice QCD 
calculations are more straightforward than for TMDs. In order to relate TMDs 
and other distributions to lattice QCD calculations there have been a number 
of novel techniques developed in recent 
years~\cite{Musch:2011er,Ji:2013dva,Lin:2014zya}.
 Comparison of precise extractions of these distributions from experimental 
data  with {\it ab initio} calculations will be very important in the future.
However, we emphasize, separation of the GPD helicity structures ($H$, $E$, $\tilde H$, $\tilde E$...) and different TMDs ($f_{1T}^\perp$, $h_1$ ...)
   is feasible only with a facility   capable of having both longitudinally and transversely polarized  beams.

One of the key applications of GPDs is the ``Ji-relation"~\cite{Ji:1997ek,*Ji:1997pf}  that allows one to relate the quark
angular momentum (spin plus orbital) to the $x$-moment of GPDs.  By subtraction, this also
allows a determination of  the quark orbital angular momentum as well as the gluon angular 
momentum.  Significant progress has been made since the EIC White Paper to better interpret the terms that appear in the nucleon spin  
decomposition~\cite{Burkardt:2008ps,Leader:2013jra} in terms of both 
twist-three quark-gluon correlations~\cite{Hatta:2011ku,Ji:2012ba},
 with also a natural description of orbital angular momentum in terms of  Wigner functions~\cite{Lorce:2011kd,Ji:2012sj}. An EIC will be crucial to drive both the theoretical and 
experimental programs to understand and explore the partonic 
content  of the nucleon spin.

Let us emphasize the requirements for EIC:
 A study of sea quark and gluon distributions for GPD and TMD programs is possible with a complete measurement of DVCS and deeply virtual meson production (DVMP) (including non diffractive pseudo scalar
production) for GPDs and SIDIS measurements with different produced hadrons ($\pi^\pm, K^\pm, p, \bar p, \ldots)$ for TMDs and spectator proton tagging on polarized light-ion beams in both cases.

Both TMDs and GPDs are extremely important for our understanding of the nucleon as a dynamical relativistic system of quarks and gluons and 
 represent future ``text book knowledge", see Fig.~\ref{fig:3D}.   These distributions are related to the orbital motion of partons inside of the nucleon and ultimately will lead to better understanding of underlying dynamics of strong color force.

\begin{figure}[t]
\begin{minipage}[t]{0.47\linewidth}
\includegraphics[width=20pc]{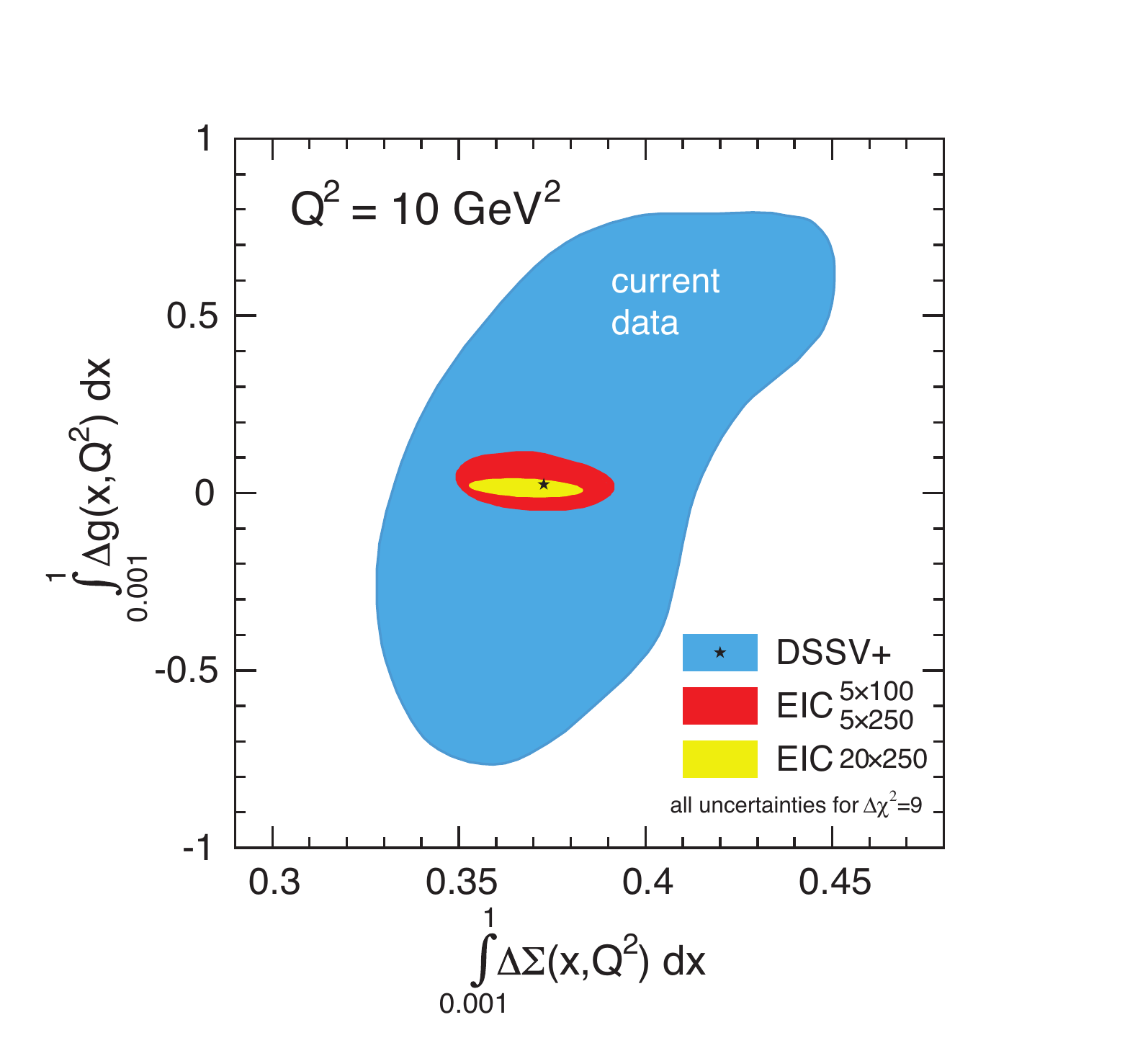}
\caption{
(Color online) 
Current knowledge of contribution from quarks $\Delta \Sigma$ and gluons $\Delta G$ coming from the region of $0.001 < x < 1$ to the spin of the nucleon compared with projected uncertainty at an EIC at two different energies. 
 Figure from Ref.~\cite{Accardi:2012qut}.
}\label{fig:DGDS} 
\end{minipage}
\hspace{1pc}%
\begin{minipage}[t]{0.46\linewidth}
\vspace{-18.7pc}%
\includegraphics[width=18pc]{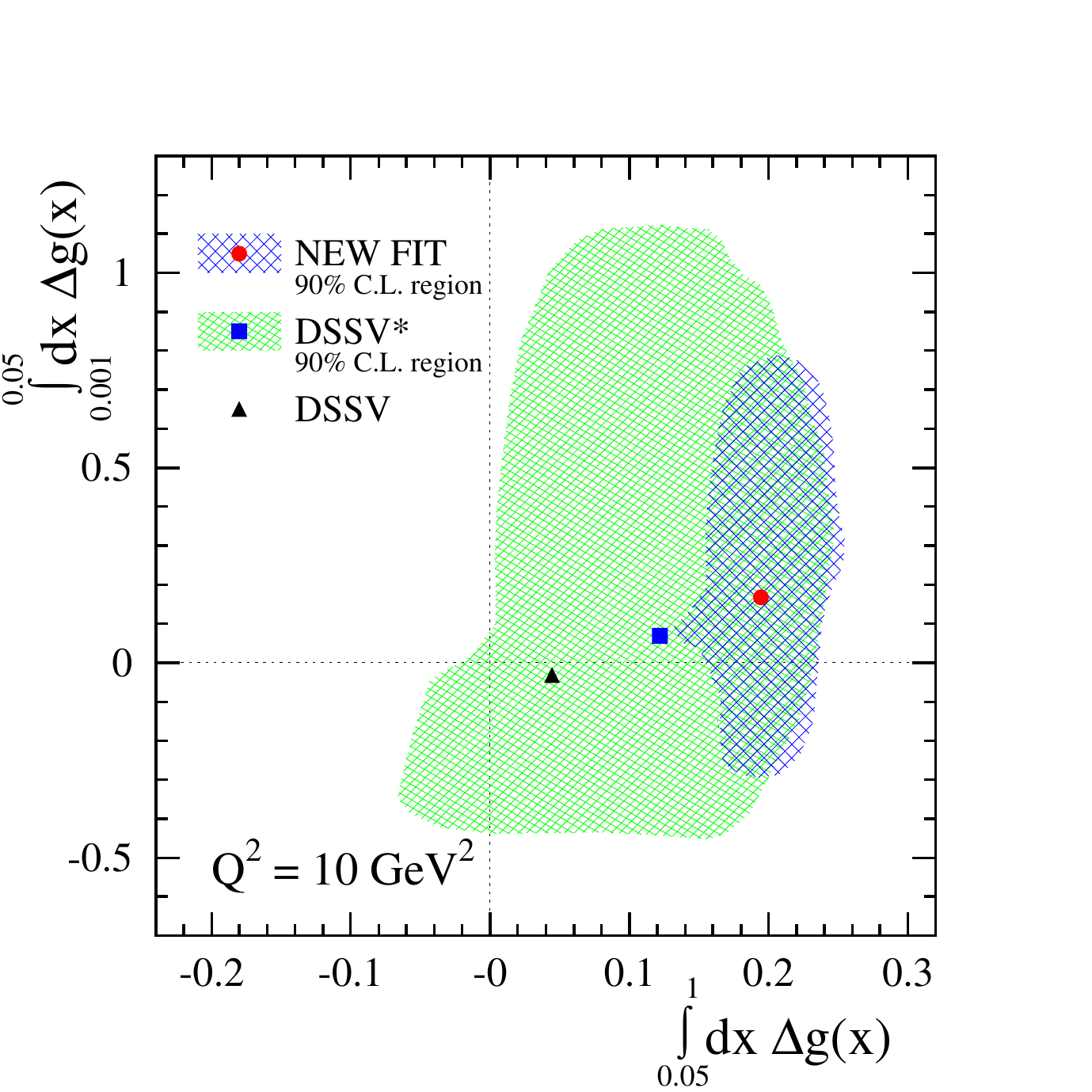}
\vspace{.5pc}%
\caption{
(Color online) 
$90\%$ C.L. areas in the plane spanned by the truncated moments of 
$\Delta g$ computed for $0.05 \leq x \leq 1$ and $0.001\leq x\leq 0.05$ at 
$Q^2 = 10 \,\mathrm{GeV}^2$ . 
Results for DSSV, DSSV*  
and our new analysis, with the symbols corresponding the respective values of each central fit, are shown.
 Figure from Ref.~\cite{deFlorian:2014yva}}\label{DSSV}
\end{minipage} 
\end{figure}

\begin{flushleft}
{\bf Helicity structure of the nucleon:}    
\end{flushleft}
The past few decades have witnessed some extraordinary progress in the
measurement and understanding of the helicity structure of the nucleon.  
The inclusive structure function $g_1(x,Q^2)$, whose unexpectedly small 
size for $x < 0.1$ revived interest in the internal spin structure of the
nucleon and gave rise to the ``nucleon spin puzzle'', 
is now measured down to $x \simeq 0.004$ for $Q^2 > 1\,\mathrm{GeV^2}$.
The RHIC spin physics program has yielded precision data on the
beam-helicity dependent production rates for inclusive jets and pions
in polarized proton--proton collisions, which in global QCD analyses
give evidence for a positive polarization of gluons in the proton for
$x \ge 0.05$.
Further advances will come from ongoing analyses of RHIC data at the
top energy of $\sqrt{s} = 510\,\mathrm{GeV}$, both for the polarization
of gluons through jet measurements and the polarization of the light
(anti)quarks through measurements of the spin-dependent rates of
leptonic decays of weak bosons produced at this energy.
They are also anticipated from proposed future running periods
and from data with new instrumentation closer to the beam regions.  
In the large-$x$ region, the 12~GeV program at Jefferson Lab is expected to determine the
valence quark helicity distributions, settling
decades-old questions about the behavior of polarized to unpolarized
PDF ratios in the $x \to 1$ limit.

Despite these advances, it is currently not possible to conclusively 
answer the basic questions on the nucleon spin composition. The total 
contribution from gluon spin  to the nucleon spin, for example, is 
uncertain to a level where it could be zero, or even larger than the nucleon spin 
itself, see Fig.\ref{fig:DGDS},  which would preclude any conclusions on possible contribution
from orbital angular momentum via the spin sum rule. 
The quark and anti-quark spin contributions are also currently limited
by uncertainties from the unmeasured region at small $x$.
However a recent 
global analysis~\cite{deFlorian:2014yva} of new experimental data on the double longitudinal
spin asymmetry in $p p$ scattering
in jet and $\pi^0$ production at RHIC~\cite{Adare:2014hsq,*Adamczyk:2014ozi}
 finds a non trivial positive gluon polarization at intermediate momentum scales. These new results
and analyses could indicate that gluons make a significant contribution
to the nucleon spin sum rule. The integrated 
 from gluon helicity  $\Delta G(Q^2)$, is displayed in Fig.\ref{DSSV} at $Q^2=10$~GeV$^2$, and the contribution to $\Delta G(Q^2)$ from the region 
covered by experimental data indicate positive gluon polarization and there is a big uncertainty coming from the region of small $x$ which
will be covered by EIC.

The proposed program with a polarized EIC brings unique capabilities
to the study of the nucleon spin. In particular, an EIC will enable
measurements that extend the kinematic range of existing and
forthcoming data to smaller values  of $x$ by between one and two 
orders of magnitude
at the same resolution scales.  At fixed values of $x$, the EIC will
allow measurements over an unprecedented range in $Q^2$.
The EIC White Paper~\cite{Accardi:2012qut} contains the impact from simulated measurements of 
$g_1(x,Q^2)$ on the (anti)quark and gluon spin contributions to the 
nucleon spin.  Besides probing deeply into the gluon-dominated region
of the nucleon, these measurements should lead to textbook knowledge
on the fundamental composition of the nucleon spin.  Semi-inclusive
measurements with one or more identified hadrons in the final state
will result in comparably impressive advances in delineating the
flavor structure of the quark and anti-quark spin 
contribution~\cite{Aschenauer:2012ve}.  
Furthermore, inclusive electro-weak measurements, utilizing both neutral and
charged currents~\cite{Adare:2010xa,*Adamczyk:2014xyw} will give access to novel structure functions and provide a unique avenue for probing the flavor decomposition
of the polarized nucleon sea~\cite{Aschenauer:2013iia}.

\begin{flushleft}
{\bf Conclusion:}
\end{flushleft}
In conclusion, an EIC will be  a precision QCD collider probing
nuclei  in the region dominated by ``sea'' quarks and gluons.
Its development 
represents an exciting opportunity 
to open the next QCD frontier for nuclear physics.  It promises to answer
fundamental outstanding questions about the structure of hadrons,
such as the decomposition of the nucleon spin,  
as well provide an opportunity to unravel the three-dimensional 
sea quark and gluon structure of visible matter.

{\bf Acknowledgments:} We thank all participants of the workshop and our colleagues for stimulating discussions. 
This work was supported in part by DOE 
contracts No.~DE-AC05-06OR23177 (J.C.,W.M.,A.P.,D.R.,C.W.), DE-AC02-98CH10886 (E.A., L.B., J.Q.), DE-AC02-05CH11231 (E.S., F.Y.), DE-AC02-76SF00515 (S.B.),  
 DE-FG02-07ER41460 (L.G.), and DE-AC52-06NA25396 (Z.K.).


\bibliography{mybiblio}

\end{document}